**Comments on "Factors associated with the spatial heterogeneity of the first wave of COVID-19 in France: a nationwide geo-epidemiological study" by Gaudart et al.**

José A. Ferreira
Department of Statistics, Informatics and Modelling
National Institute for Public Health and the Environment (RIVM)
Antonie van Leeuwenhoeklaan 9,
3721 MA Bilthoven
The Netherlands
e-mail: jose.ferreira@rivm.nl

**Abstract**
In a recent paper, Jean Gaudart and colleagues studied the factors associated with the spatial heterogeneity of the first wave of COVID-19 in France. We make some critical comments on their work which may be useful for future, similar studies.

Explaining the ravages of the covid-19 epidemic and the greater or lesser success in combating it will be of interest in months and probably years to come. The study of Gaudart and colleagues[1] represents a worthy effort to identify factors that aggravate mortality due to the epidemic, but we have some critical comments on it which may be useful for similar studies. The first one concerns the statistical methods: What makes things 'scientific' (irrespectively of their being correct or not) is not the collecting of data, not the formulating of hypotheses, not the use of computation, but the reasoning in them. Although the models and methods used in the study are enumerated, there is little or no reasoning behind them; we doubt that there could be any, but it is the burden of justifying the models (of showing that they correspond, in some sense, with the data) that lies with the authors rather than the burden of disproving them that lies with the readers. It is true that the authors follow common practice, but that is no excuse because our objections are well known:[2,3,4,5] in particular, p-values may be small for no reason other than that the models under which they are computed are wrong, and despite sensitivity and analyses of residuals to confirm that they are approximately right; and to the best of our knowledge the software 'dagitty' is meant to identify sets of variables the conditioning upon which permits the estimation of a causal effect, but this presupposes a model for the relationships of cause and effect between variables, while, for example, in the model represented by graph A of the appendix 'economic indicators' can hardly be thought to cause 'population age structure' (surely there are causal relations between these two, but they are 'dynamic' and far too complicated to be compressed into a single, one-sided relation). Classical 'model building' techniques for arriving at approximately correct models and for defending them could be useful, but they are exacting and are hardly applicable to situations involving more than just a few variables.[6] Older methods involving stratification on 'background' variables combined with permutation versions of non-parametric tests avoid or greatly reduce biases in p-values, but the associations they yield normally retain some confounding (their use implying no pretension to remove confounding to the point of exhibiting 'pure' or causal relations, which in any case is impossible in most observational studies), and we are not sure that they could have served the analyses of Gaudart and colleagues which are based on 90 to 100 sampling units (the French departments). Despite this relatively small sample size, perhaps statistical prediction analyses based on a non-parametric algorithm would provide clear information about which variables are most useful to predict incidence, mortality and fatality rates, and about how accurate the predictions of these outcomes are. Used in this study, analyses of this kind (which derive evidence from various estimates of prediction accuracy rather than from p-values) would yield predictions at each location based on 'training data' consisting of the outcomes and predictor variables (the regional centroid, economic indicators, etc.) from the other locations, and the accuracy of the predictions could be quantified overall and geographically. In particular, random forest[5] might give reasonably accurate predictions, and with them a meaningful ranking of the variables according to their predictive value, which is not so readily or accurately obtainable from the models fitted by the authors. (In principle, prediction analyses could be carried out with the models used in the paper, but a cross-validation procedure properly accounting for the 'model building' would require more data.)

It is unreasonable to expect evidence for clear-cut causal effects in an observational study like this (studies which provide that being exceptional), and finding associations is a perfectly legitimate and most useful goal, certainly if the findings can be given some sort of causal explanation. Despite the criticism vented above, the associations presented by the authors seem genuine and to have been given mostly correct (though necessarily partial) causal interpretations when appropriate. However, a couple of important statements in the paper seem inaccurate to us and certain of its conclusions are not correct or at least not warranted by the study, which brings us to the second part of our comment.

It is said that "Case fatality rate was not associated with the initial number of intensive care beds, suggesting that hospitals could rapidly scale up their capacities and organise medical evacuations to less affected areas." The authors mean to say that they have found no evidence for the association, and yet interpret the non-rejection of a null hypothesis by a positive statement about how events turned out. We think that such a suggestion requires further qualification, and that the events it refers to are best described by collecting and analyzing testimonies of those involved in them, as in a historical investigation. In this connection it is also said that "the number of ICU beds available in 2018 was not associated with mortality or case fatality rate", but the association in question may exist and escape detection. And while the statement that "we found no association between baseline population health and health-care services and incidence and mortality rates" offers no doubts, the authors revert to a positive, seemingly unwarranted statement: "This finding suggests that hospitals managed to scale up their ICU capacity (≥100% increase in 21 departments; appendix p 19) or organise medical evacuations to less affected departments when necessary."

The authors say that the time "between the first COVID-19-associated death and the onset of the lockdown appeared to be positively associated with in-hospital incidence, mortality, and case fatality rates". This is alright, but as far as we can see it need not imply anything about the lockdown: the interval of time seems to indicate how early the infection had been around. So the following seems unwarranted: "In other words, morbidity and mortality were lower in departments where the general lockdown caught the epidemic at an earlier stage of its expansion. This finding suggests that the lockdown was an effective way to control the diffusion of this wave of the epidemic across the country." If the interval had been defined with respect to an event other than the lockdown then the association would still have been detected. As to a 'pure' effect of the lockdown (supposedly separate from other effects, such as the general alarm or the self-imposed effort of many families to protect their own elderly), we wonder whether this can ever be determined; at any rate, and although it is hard to imagine that the lockdown had no consequences in France, it seems well to report that even now there are studies, also based on statistical parametric models, which draw different conclusions about those consequences.[7] Incidentally, the authors say that "our ecological study was not designed to assess the effectiveness of chloroquine and hydroxychloroquine against

COVID-19 at the individual level; these drugs have been shown not to be effective", so it may be well to add that on this subject too there continue to be different opinions.[8]

The authors end the paper by saying that "In conclusion, our findings outline the effect of the COVID-19 pandemic wave in a country that could absorb the shock, thanks to a strong hospital system and a national lockdown. However, the findings indirectly underscore the weakness of its preventive and public health system, which could be useful for informing countries' preparedness for the current or future pandemic waves." The first statement appears to include a value judgement and an attribution of cause, neither of which is warranted by the study; the second is similar in nature but seems to be contrary in meaning.

Our last point is more of a recommendation since it concerns factors which, probably because data on them were not available at the time, have not been considered in this study and which in our view could be important in future studies. Evidence from East Asia[9,10,11,12] suggests that early treatment, isolation and accompaniment of covid-19 patients, even of those not, or not yet, showing serious symptoms, is important for reducing mortality and transmission, and accordingly, already by April 2020, scientists and practitioners, such as Dr. X. Pothet, urged authorities in France to take heed of that evidence[12,13]. Despite possible social and climatic particularities of the country, the apparent success of Singapore in dealing with the epidemic supports the idea that treating patients as early as possible in or out of hospital instead of asking them to quarantine as long as their symptoms are mild, is an effective way of reducing mortality:

> "The majority of patients picked up by our testing have mild or no symptoms. Such patients are generally admitted to a Community Care Facility (CCF) where most recover with minimal intervention. Patients in these facilities are monitored closely in case they need to be transferred to hospital for better management and support.";[14,15]

> "Thus, triaging becomes very important. The duration from the initial symptoms of COVID-19 to respiratory failure in most patients is ~ 7 days […]. Many patients go on to develop 'silent hypoxemia', so-called because of its insidious and hard-to-detect nature. It has been reported that unlike pneumonia due to other infections, COVID-19 pneumonia patients may not feel dyspnoeic or any noticeable discomfort in chest. The physical manifestations become evident when pneumonia has deteriorated to moderate-to-severe levels"; [16]

> "Since the chain of events triggered by SARS-CoV-2 infection evolves quickly, any planned intervention must come as early as possible. Besides, since the pathogenesis of COVID-19 involves non-viral mechanisms, any intervention planned must also address the correction or modulation of these disbalances. Hence, any therapeutic intervention must be early and combine antiviral and adjuvant therapies. However, the moment of diagnosis and eventual hospital admission will mark the timeframe of interventions"[11].

So it may be that, in addition to age, comorbidities and the time since the beginning of the epidemic, the best predictors of mortality and fatality rate are the type of treatment and the stage of the disease at which treatment is provided, to the extent that the inclusion of these variables in prediction analyses will reveal those outcomes to be quite predictable. As Gaudart and colleagues point out, the criteria for admission to hospital varied across departments, and the provision of treatment in general, in or out of hospital, must have varied as well, in space and also in time, and so too must have varied the protocols for treating covid-19 patients. Thus, these variations, not only within France but between countries, should offer opportunities for identifying the most important contributors to mortality due to covid-19. Of course, the reliability of data on the factors in question is important, as is the resolution at which they or their substitutes are available; but there ought to be records of, and testimonies about, the criteria for the provision of treatment and the protocols adopted by the various health care institutions for treating patients during the epidemic, from which suitable variables can be distilled, and at a more detailed level than the departmental one. A recognition of the importance of the timely provision of treatment of covid-19 might even turn out to extend to other infectious diseases, which would strengthen current calls for a 'paradigm shift' in infection management.[17]